\documentclass[showpacs,aps,floatfix,pra,reprint,superscriptaddress]{revtex4-1}
\usepackage{footnote}
\makesavenoteenv{tabular}
\usepackage{scrextend}
\usepackage{xfrac}
\usepackage{amssymb}
\usepackage{braket}
\usepackage{graphicx}
\usepackage{verbatim}
\usepackage{mhchem}
\usepackage{etoolbox}\usepackage{amsfonts} 
\usepackage{amsmath} 
\usepackage[utf8]{inputenc} 
\usepackage{mathtools}
\usepackage{soul,xcolor,colortbl}
\usepackage{hyperref} \hypersetup{colorlinks=true,citecolor=blue,linkcolor=blue,urlcolor=blue} 
\usepackage{tabularx}
\usepackage{bm} 
\usepackage{array}
\usepackage{color} 
\newcolumntype{L}[1]{>{\raggedright\let\newline\\\arraybackslash\hspace{0pt}}m{#1}}
\newcolumntype{C}[1]{>{\centering\let\newline\\\arraybackslash\hspace{0pt}}m{#1}}
\newcolumntype{R}[1]{>{\raggedleft\let\newline\\\arraybackslash\hspace{0pt}}m{#1}}
\def\ACBN0{{\small ACBN0}}
\def\APL{{\small APL}}
\def\AAPL{{\small AAPL}}
\def\AGL{{\small AGL}}
\def\APL{{\small APL}}
\def\QHA{{\small QHA}}
\def\QHAAPL{{\small QHA-APL}}
\def\AEL{{\small AEL}}
\def\AFLOW{{\small AFLOW}}
\def\AFLOWPI{{{\small AFLOW}$\pi$}}
\def\VASP{{\small VASP}}
\def\GIBBS{{\small GIBBS}}
 
\def\BTE{{\small BTE}}
\def\IFC{{\small IFC}}
\def\IFCs{{\small IFC}s}
\def\LDA{{\small LDA}}
\def\GGA{{\small GGA}}
\def\PBE{{\small PBE}}

\def\RMSRD{{\small RMSrD}}

\def\citeAFLOWAGL{\cite{curtarolo:art96, curtarolo:art115}}
\def\citeAFLOW{\cite{aflowPAPER,aflowlibPAPER,aflowBZ,curtarolo:art92,Calderon_cms_2015,curtarolo:art63,curtarolo:art57,curtarolo:art49,monsterPGM}}

\def\sAAPL{{\substack{\scalebox{0.6}{AAPL}}}}

\def\sexp{{\substack{\scalebox{0.6}{Exp}}}}

\begin{document}
\title{Predicting the lattice thermal conductivity of solids by solving the Boltzmann transport equation: AFLOW - AAPL an automated, accurate and efficient framework}.

\author{Jose J. Plata} \affiliation{Department of Mechanical Engineering and Materials Science, Duke University, Durham, North Carolina 27708, USA}
\author{Demet Usanmaz} \affiliation{Department of Mechanical Engineering and Materials Science, Duke University, Durham, North Carolina 27708, USA}
\author{Pinku Nath} \affiliation{Department of Mechanical Engineering and Materials Science, Duke University, Durham, North Carolina 27708, USA}
\author{Cormac Toher} \affiliation{Department of Mechanical Engineering and Materials Science, Duke University, Durham, North Carolina 27708, USA}
\author{Jesus Carrete} \affiliation{CEA-Grenoble, 17 Rue des Martyrs, Grenoble 38000, France}
\author{Mark Asta} \affiliation{Department of Materials Science and Engineering, University of California, Berkeley, 210 Hearst Memorial Mining Building, Berkeley, USA}
\author{Maarten de Jong} \affiliation{Department of Materials Science and Engineering, University of California, Berkeley, 210 Hearst Memorial Mining Building, Berkeley, USA}
\author{Marco Buongiorno Nardelli} \affiliation{Department of Physics and Department of Chemistry, University of North Texas, Denton TX, USA.}
\author{Marco Fornari} \affiliation{Department of Physics, Central Michigan University, Mount Pleasant, MI 48858, USA.}
\author{Stefano Curtarolo} \affiliation{Materials Science, Electrical Engineering, Physics and Chemistry, Duke University, Durham NC, 27708} \email[]{stefano@duke.edu}

\date{\today}

\begin{abstract}
  One of the most accurate approaches for calculating lattice thermal conductivity, $\kappa_l$, is solving the Boltzmann transport equation starting from third-order anharmonic force constants.
  In addition to the underlying approximations of {\it ab-initio} parameterization, two main challenges are associated with this path.
  High computational costs and lack of automation in the frameworks using this methodology
  affect the discovery rate of novel materials with {\it ad-hoc } properties.
  Here, we present the \underline{A}utomatic-\underline{A}nharmonic-\underline{P}honon-\underline{L}ibrary, \AAPL.
  It efficiently computes interatomic force constants by making effective use of crystal symmetry analysis, 
  it solves the Boltzmann transport equation to obtain $\kappa_{l}$,
  and allows a fully integrated operation with minimum user intervention, a rational addition to the current high-throughput accelerated materials development framework \AFLOW.
  We show an ``experiment {\it versus} theory'' study of the approach, we compare accuracy and speed with respect to other available packages, and for materials 
  characterized by strong electron localization and correlation, we demonstrate that it is possible to improve accuracy without increasing computational requirements by 
  combining \AAPL\ with the pseudo-hybrid functional \ACBN0.
\end{abstract}
\pacs{63.20-e, 63.20.kg, 66.70.-f}

\maketitle

\section{Introduction}

Lattice thermal conductivity, $\kappa_{l}$, is the key materials' property for many technologies and applications such as 
thermoelectricity \cite{zebarjadi_perspectives_2012,curtarolo:art84}, heat sink materials 
\cite{Yeh_2002}, rewritable density scanning-probe phase-change memories \cite{Wright_tnano_2011} and
thermal medical devices \cite{Cahill_APR_2014}.
Fast and robust predictions of this quantity remain a challenge \cite{Pinku_sm_2016}: 
semi-empirical models \cite{Ziman_1960,callaway_model_1959,Allen_PHMB_1994} 
are computationally inexpensive but require some experimental data.
Similarly, classical molecular dynamics combined with Green-Kubo relations \cite{Green_JCP_1954,Kubo_JPSJ_1957,curtarolo:art12} is
reasonably quick but requires the knowledge of specific force fields. 
On the contrary, frameworks based on the quasiharmonic Debye model, such
as \GIBBS\ \cite{Blanco_CPC_GIBBS_2004} or the 
\underline{A}FLOW-\underline{G}ibbs-\underline{L}ibrary, 
\AGL\ \citeAFLOWAGL, are extremely
efficient as pre-screening techniques but they lack quantitative
accuracy.

The \underline{q}uasi\underline{h}armonic \underline{a}pproximation, \QHA, alone has also been used in different models to predict $\kappa_{l}$ \cite{Madsen_PRB_2014, Pinku_sm_2016}. 
Although QHA based models overall improve accuracy of $\kappa_{l}$, they are far 
from the results obtained from calculating the anharmonic force constants and solving the
associated \underline{B}oltzmann \underline{t}ransport \underline{e}quation, \BTE\ \cite{Ziman_1960, mingo_ab_2014}. 
To the best of our knowledge, solving the BTE is the best method for systematically and accurately calculating thermal conductivity \cite{Broido_APL_231922_2007, ward_intrinsic_2010, Tang_pnas_2010}. 
This approach has been successfully applied to many systems during the last decade.
It has been recently implemented in packages including Phono3py \cite{Togo_prb_2015}, 
PhonTS \cite{Chernatynskiy_cpc_2015}, {\small ALAMODE} \cite{Tadano_jpcm_2014},
and ShengBTE \cite{ShengBTE_2014}, which compute $\kappa_{l}$ by calculating 
the anharmonic force constants and solving the \BTE. 
Nevertheless, there is a lack of a robust framework, able to calculate
$\kappa_{l}$ with minimum intervention from the user and 
therefore targeted to high-throughput automatic and accelerated
materials discovery.

Many challenges need to be tackled.
{\bf I.}
The third order \underline{i}nteratomic \underline{f}orce
\underline{c}onstants (\IFCs) up to a certain distance cut-off 
are computationally expensive to obtain from first principles.
Overall they represent the major concern for the method.
Effective use of crystalline symmetry of the system must be
employed to map, through appropriate tensorial transformations, dependent \IFCs\
and therefore reduce the number of calculations. The task is performed
by the internal \AFLOW\ point-factor-space group calculator \cite{curtarolo:art121}.
Recently, it has also been proposed to obtain the \IFCs\ 
by inverting the results of many entangled calculations
with the use of compressive sensing \cite{Fei_PRL_2014}. Further
studies need to be carried out to address the scaling of the algorithm
with respect to cut-offs and accuracy.
{\bf II.}
For a rational software for accelerated materials development, all the geometric optimizations, symmetry analyses, supercell creation,
pre and post-processing, and automatic error corrections to get the \IFCs\ in addition to the appropriate integration for the \BTE\ must be
performed by a single code.
Here, we present \AAPL, which computes the \IFCs\ and solves the \BTE\ to predict $\kappa_{l}$ as part of the \AFLOW\ high-throughput
framework \citeAFLOW, automatizing the entire process. 
The software is being finalized for an official open-source release during 2017,
within the {\small GNU GPL} license and, in addition to the \AFLOW\ consortium framework, it will also be implemented inside the Materials Project Ecosystem \cite{APL_Mater_Jain2013}.
{\bf III.}
The accuracy of the method mostly depends on the accuracy of the computed forces, and therefore it will inherit the same limitations as
the {\it ab-initio} method used.
For materials characterized by strong electron localization and correlation, accurate hybrid functionals for Density Functional Theory 
parameterizations might not even be feasible as they would drastically increase computational costs, with respect to more basic
\LDA\ or \GGA\ functionals.  
In that case, new strategies should be developed to contain computational demands. 
Here we give an example: we demonstrate that it is possible to improve the accuracy without increasing computational requirements 
by combining \AAPL\ with the pseudo-hybrid functional \ACBN0\
\cite{curtarolo:art93,curtarolo:art116,curtarolo:art111,curtarolo:art108,curtarolo:art105,curtarolo:art103,curtarolo:art86}.

\section{Methodology: The \underline{A}utomatic-\underline{A}nharmonic-\underline{P}honon-\underline{L}ibrary (\AAPL) }

{\bf The Boltzmann transport equation.}
The Boltzmann equation for phonons, originally formulated by Peierls in 1929, is an important approach for studying phonon transport \cite{Ziman_1960}. 
Its solution has posed a challenge for the last several decades.
Callaway \cite{callaway_model_1959} and Allen \cite{Allen_prb_2013} proposed models based on parameters that are fitted to experimental data.
In 2003, Deinzer \textit{et al}. used \underline{d}ensity
\underline{f}unctional \underline{p}erturbation \underline{t}heory
({\small DFPT})
to study the phonon linewidths of Si and Ge \cite{Deinzer_PRB_2003}.
Since then, many authors have used the solution of the \BTE\ to calculate the lattice thermal conductivity of solids \cite{Broido_APL_231922_2007, ward_intrinsic_2010, Tang_pnas_2010}.
The most used approach is the iterative solution of the \BTE\ proposed
by Omini \textit{et al}. and successfully applied in the prediction of
the $\kappa_{l}$ tensor for different materials \cite{Omini_pscb_1995, Omini_prb_1996, omini_heat_1997}: 
\begin{equation}
  \kappa^{\alpha\beta}_{l} = \frac{1}{N \Omega k_{\mathrm{B}} T^2}
  \sum_{\lambda} f_0 (f_0 + 1) (\hbar \omega_{\lambda})^2
  {v^{\alpha}_{\lambda}} F^{\beta}_{\lambda},
  \label{equationKappa}
\end{equation}
where superscripts $\alpha$ and $\beta$ are two of the Cartesian
direction indices  and 
the subscript $\lambda$ comprises both phonon branch index $i$ and a wave vector {$\mathbf{q}$}. 
The variables $\omega_{\lambda}$ and $\mathbf{v}_{\lambda}$ are the angular frequency and group velocity of the phonon mode 
$\lambda$ respectively, while $f_0 (\omega_{\lambda})$ is the phonon distribution function according to 
Bose-Einstein statistics. 
%
All these quantities are obtained through the calculation of the
\IFCs\ by using a finite-difference supercell approach: forces {\it versus} small
displacement of inequivalent atoms.
In this approach, a reference unit cell of volume $\Omega$ is used to create
the supercell up to the cut-off distance.
For the various summations, the Brillouin zone, BZ, is discretized into a
\text{$\Gamma$}-centered orthogonal regular grid of 
$N \equiv N_1 \times N_2\times N_3$ $\mathbf{q}$-points, where subscripts $1$, $2$, and $3$
indicate the lattice vector indices. 

The mean free displacement $\mathbf{F}_{\lambda}$ follows the definition of
the Bose-Einstein phonon distribution, $f_\lambda$,  in the presence of a temperature gradient $\nabla T$.
For small perturbations,  $\nabla T\sim 0$, $f_\lambda$ can be
expanded as $f_\lambda \sim f_0 (\omega_{\lambda}) + g_\lambda$, 
where $g_\lambda$ is the first-order non-equilibrium contribution
linear in $\nabla T$:
\begin{equation}
  g_\lambda \equiv -\mathbf{F}_{\lambda}\cdot\nabla T \frac{df_0}{dT}.
  \nonumber
\end{equation}
Finally, the BTE can be expressed as a linear system of equations for $\mathbf{F}_{\lambda}$, as \cite{Omini_pscb_1995,omini_heat_1997,Omini_prb_1996,ward_ab_2009,ward_intrinsic_2010,Lindsay_JPCM_2008}:

\begin{eqnarray}
  \mathbf{F}_{\lambda}&=&\tau_{\lambda}^0(\mathbf{v}_{\lambda}+\boldsymbol{\Delta}_{\lambda}) \label{equationF} \\
     \mathbf{\Delta}_{\lambda}&=& \frac{1}{N}\Bigg(\sum\limits_{\lambda'\lambda''}\limits^{+}\Gamma^{+}_{\lambda\lambda'\lambda''}\big(\xi_{\lambda\lambda''} \mathbf{F}_{\lambda''}-\xi_{\lambda\lambda'}\mathbf{F}_{\lambda'}\big) + \nonumber\\
    && + \sum\limits_{\lambda'\lambda''}\limits^{-} \frac{1}{2}\Gamma^{-}_{\lambda\lambda'\lambda''}\big(\xi_{\lambda\lambda''} \mathbf{F}_{\lambda''}+\xi_{\lambda\lambda'}\mathbf{F}_{\lambda'}\big) +\nonumber\\
    && + \sum\limits_{\lambda'}\Gamma_{\lambda\lambda'}\xi_{\lambda\lambda'} \mathbf{F}_{\lambda'}\Bigg),\nonumber
\end{eqnarray}
with $\xi_{\lambda\lambda'}= \omega_{\lambda}/\omega_{\lambda'}$. 
The frequently used \underline{r}elaxation \underline{t}ime
\underline{a}pproximation, {\small RTA}, corresponds to neglecting the
$\boldsymbol{\Delta}_{\lambda}$ correction. For the fully 
solution, $\mathbf{F}_{\lambda}$ can be  self-consistently
solved starting from the {\small RTA} guess, until convergence of
$\kappa_{l}$, Eq. (\ref{equationKappa}). 
The other quantities present in these formulas, the relaxation time $\tau_{\lambda}^0$, and the three-phonon 
scattering rates $\Gamma^{\pm}_{\lambda\lambda'\lambda''}$, will be illustrated in the next Section. 

{\bf Vibrational Modes and group velocities.}
The vibrational modes are obtained  by diagonalizing the dynamical matrix 
$D(\mathbf{q})$ \cite{Ziman_1960,ThermoCrys,IntroLattDym,PhysPhon}:
\begin{equation}
  D(\mathbf{q})\mathbf{e}_{\lambda}=\omega^2_{\lambda}\mathbf{e}_{\lambda};
  \label{eq:dynamical1}
\end{equation}
\begin{equation}
  D_{ij}^{\alpha \beta }(\mathbf{q}) = \sum_l \frac{\Phi(i,j)_{\alpha \beta}}{\sqrt{M(i)M(j)}}\exp\left[-i\mathbf{q} \cdot \left(\mathbf{R}_{l}-\mathbf{R}_{0}\right)\right],
  \label{eq:dynamical2}
\end{equation}
where $M(j)$ is the mass of $j$-atom, $\mathbf{e}_{\lambda}$ is the eigenvector for $\lambda$, $\mathbf{R}_{l}$ is 
the position of lattice point $l$ and $\Phi_{ij}^{\alpha \beta}$ are the second-order force constants.
$D(\mathbf{q})$ is a Hermitian $3n_a\times3n_a$ matrix, where the
factor ``3'' comes from the dimensionality of the problem,
and $n_a$ represents the number of atoms in the unit cell. 

The non-analytical contributions to the dynamical matrix are included
by using the formulation of Wang \textit{et al.} \cite{Wang2010}:
\begin{equation}
  \widetilde{D}_{ij}^{\alpha \beta }(\mathbf{q}) = \frac{4\pi
    e^2}{\Omega}\frac{\left[\mathbf{q}\cdot\mathbf{Z}(i)\right]_{\alpha}\left[\mathbf{q}\cdot\mathbf{Z}(j)\right]_{\beta}}{\mathbf{q}\cdot\epsilon_{\infty}\cdot\mathbf{q}}\exp\left[-i\mathbf{q} \cdot \left(\mathbf{R}_{l}-\mathbf{R}_{0}\right)\right].
  \label{eq:dynamical3}
\end{equation}
This contribution requires the calculation of the Born effective charge tensors, $\mathbf{Z}$, and the high frequency static dielectric tensor, $\epsilon_{\infty}$, 
i.e. the contribution to the dielectric permitivity tensor from the electronic polarization \cite{BaroniRMP2001}.
Materials with high $\mathbf{Z}$ and low $\epsilon_{\infty}$ are the cases in which the non-analytical contributions
are crucial for appropriate description of the phonon spectra as they cause the
LO–TO splitting of the spectrum (between longitudinal and transverse optical phonon frequencies) \cite{BaroniRMP2001}.

The group velocities, $\mathbf{v}_{\lambda}$, follow the Hellmann-Feynman theorem:
\begin{equation}
  \mathbf{v}_{\lambda}=\frac{1}{2\omega_{\lambda}}\left\langle
    \mathbf{e}_{\lambda} \left |\frac{\partial
        D(\mathbf{q})}{\partial\mathbf{q}} \right |
    \mathbf{e}_{\lambda} \right \rangle.
  \label{eq:group_velocities}
\end{equation}

{\bf Scattering time.}
The total scattering time is a sum of terms representing different phenomena:
\begin{equation}
  \frac{1}{\tau^0_{\lambda}}=\frac{1}{\tau^{\mathrm{anh}}_{\lambda}} + \frac{1}{\tau^{\mathrm{iso}}_{\lambda}}+\frac{1}{\tau^{\mathrm{bnd}}_{\lambda}}.
    \label{eq:tau_zero}
\end{equation}

$\tau^{\mathrm{iso}}_{\lambda}$ indicates the isotopic or elastic
scattering time and it is due to the isotopic disorder \cite{tamura_isotope_1983,kundu_role_2011}:
\begin{equation}
\begin{split}
  \frac{1}{\tau^{\mathrm{iso}}_{\lambda}}&=\frac{1}{N}\sum\limits_{\lambda'}\Gamma_{\lambda\lambda'}\\
                                         &=\frac{1}{N}\sum\limits_{\lambda'}\frac{\pi\omega_{\lambda}^2}{2}\sum\limits_{i}g(i)|\mathbf{e}^*_{\lambda}(i)\mathbf{e}_{\lambda'}(i)|^2\delta(\omega_{\lambda}-\omega_{\lambda'}),
\end{split}
  \label{eq:tau_iso}
\end{equation}
where $g(i) = \sum_sf_s(i)\left[1-M(i)^{s}/\overline{M}(i)^{s}\right]^2$ is the Pearson deviation coefficient of masses $M(i)^{s}$ of 
isotopes $s$ for the  $i$ atom, $f_s$ is the relative fraction of isotope $s$, and $\overline{M}(i)^{s}$ is the average mass of the element \cite{Berglund_pac_2011}.

$\tau^{\mathrm{bnd}}_{\lambda}$ is the time associated with
scattering at the grain boundaries \cite{Wang_apl_2011, Carrete_jap_2015},
\begin{equation}
  \frac{1}{\tau^{\mathrm{bnd}}_{\lambda}}=\frac{|\mathbf{v}_{\lambda}|}{L},
  \label{eq:tau_bnd}
\end{equation}
where $L$ is the average grain size.
The effect of the boundaries on $\kappa_l$ has also been  calculated  by restricting the summation
to the modes with a mean free path, $\Lambda = \mathbf{F}_{\lambda} \cdot \mathbf{v}_{\lambda} / \left |\mathbf{v}_{\lambda} \right|$, shorter than $L$ \cite{ShengBTE_2014}:

\begin{equation}
  \kappa^{\alpha\beta}_{l,(\Lambda<L)} = \frac{1}{N \Omega k_B T^2} \sum_{\lambda}^{\Lambda_{\lambda} < L} f_0 (f_0 + 1) (\hbar \omega_{\lambda})^2 {v^{\alpha}_{\lambda}} F^{\beta}_{\lambda}.
  \label{eq:cumulative_kappa}
\end{equation}

$\tau^{\mathrm{anh}}_{\lambda}$  is the three-phonon scattering time.
It is the largest contribution to $\tau^{0}_{\lambda}$ for single
crystals at medium-temperature ranges
and it is the most computationally expensive quantity to obtain:
\begin{equation}
  \frac{1}{\tau^{\mathrm{anh}}_{\lambda}}=\frac{1}{N}\Bigg(\sum\limits_{\lambda'\lambda''}\limits^{+}\Gamma^{+}_{\lambda\lambda'\lambda''}+\sum\limits_{\lambda'\lambda''}\limits^{-}\frac{1}{2}\Gamma^{-}_{\lambda\lambda'\lambda''}\Bigg).
  \label{eq:tau_anh}
\end{equation}
Conservation of the quasi-momentum requires that $\mathbf{q}''=\mathbf{q} \pm \mathbf{q}'+\mathbf{Q}$ in the summation $\sum^{\pm}$, for 
some reciprocal lattice vector $\mathbf{Q}$ such that $\mathbf{q}''$ is in the same image of the Brillouin zone as $\mathbf{q}$ and $\mathbf{q}'$. 
The three-phonon scattering rates, $\Gamma^{\pm}_{\lambda\lambda'\lambda''}$, are computed as
\begin{equation}
  \Gamma^{+}_{\lambda\lambda'\lambda''}
  \equiv
  \frac{\hbar\pi}{4}\frac{f'_0-f''_0}{\omega_{\lambda}\omega_{\lambda'}\omega_{\lambda''}}\big|V^{+}_{\lambda\lambda'\lambda''}\big|^2\delta(\omega_{\lambda}+\omega_{\lambda'}-\omega_{\lambda''}),
\label{gamma2eq}
\end{equation}
and
\begin{equation}
  \Gamma^{-}_{\lambda\lambda'\lambda''} 
  \equiv
  \frac{\hbar\pi}{4}\frac{f'_0+f''_0+1}{\omega_{\lambda}\omega_{\lambda'}\omega_{\lambda''}}\big|V^{-}_{\lambda\lambda'\lambda''}\big|^2\delta(\omega_{\lambda}-\omega_{\lambda'}-\omega_{\lambda''}).
\label{gamma3eq}
\end{equation}
The scattering matrix elements, $V^{\pm}_{\lambda\lambda'\lambda''}$, are given by \cite{Lindsay_JPCM_2008,ward_ab_2009}
\begin{equation}
  V^{\pm}_{\lambda\lambda'\lambda''}\!\!=\!\!\!\!\!\!\sum\limits_{\substack{i\in  {\rm uc} \\ \left\{j,k\right\} \in {\rm sc}\\ \alpha\beta\gamma}}\!\!\!
  \Phi(i,j,k)_{\alpha \beta \gamma} \frac{e^{\alpha}_{\lambda}(i)e^{\beta}_{p',\pm
      \mathbf{q'}}(j)e^{\gamma}_{p',-
      \mathbf{q'}}(k)}{\sqrt{M(i)M(j)M(k)}},
  \label{eq:scattering_V}
\end{equation}
where $\Phi(i,j,k)_{\alpha \beta \gamma}$ are the anharmonic force
constants (introduced below) and $e^{\beta}_{p',\pm \mathbf{q'}}(j)$ is the element of
the eigenvector 
of branch $p'$ at point $\pm \mathbf{q'}$ that corresponds to $j$-atom in the $\beta$-direction.
Note the indices $\{i \in {\rm uc\ (unit\ cell)\}}$ while $\{j,k\in {\rm sc\ (supercell)}\}$.
The conservation of energy, enforced by the Dirac distribution, can
cause numerical instability during the calculations. Thus, we follow Li
\emph{et al.} \cite{ShengBTE_2014} and substitute $\delta$ with a
normalized Gaussian distribution $g$: $\delta(\cdots)\rightarrow
g(\cdots)$ in Eqns. (\ref{gamma2eq}-\ref{gamma3eq}) with
\begin{eqnarray}
  &&g(\omega_{\lambda}-(\pm\omega_{\lambda'}+\omega_{\lambda''}))
     \equiv \frac{1}{\sqrt{2\pi}\sigma}e^{\frac{(\omega_{\lambda}-(\pm\omega_{\lambda'}+\omega_{\lambda''}))^2}{2\sigma^2}},\nonumber
  \\
  &&\sigma \equiv
  \zeta\sigma_{(\pm\omega_{\lambda'}+\omega_{\lambda''})} = \nonumber \\
  && \ \ \ =\frac{\zeta}{\sqrt{12}}\sqrt{\sum_{\nu}\left[\sum_{\alpha}\left(v^{\alpha}_{\lambda'}-v^{\alpha}_{\lambda''}\right)\frac{Q_{\nu}^{\alpha}}{N_{\nu}}\right]^2},
     \label{delta2gauss}
\end{eqnarray}
where $Q_{\nu}^{\alpha}$ is the component in the Cartesian direction, $\alpha$, of the reciprocal-space lattice vector $\mathbf{Q}_{\nu}$
and $N_{\nu}$ is the number of points of the $\mathbf{q}$-points grid in the reciprocal-space direction $\nu$ . 
In principle, the parameter $\zeta$ could be taken equal to one.
However, it can be adjusted to lower values to increase the speed of the calculations, without much effect on the overall accuracy of the integrations.

{\bf Interatomic Force Constants (\IFCs).}
The $n^{\mathrm{th}}$-order interatomic force constants \IFCs,
$\Phi(i,j,\cdots)^{\alpha \beta \cdots}$ are tensorial
quantities representing derivatives of the potential energy ($V$)
with respect to the atomic displacements from equilibrium:
\begin{equation}
  \begin{split}
    V &= V_0 + \frac{1}{2!} \sum_{ij, \alpha\beta} \Phi(i,j)_{\alpha \beta} r(i)^{\alpha} r(j)^{\beta} + \\
    & +   \frac{1}{3!} \sum_{ijk, \alpha\beta\gamma} \Phi(i,j,k)_{\alpha \beta \gamma} r(i)^{\alpha} r(j)^{\beta} r(k)^{\gamma}  + \cdots
  \end{split}
  \label{eq:potential}
\end{equation}
Labels $i, j, k, \cdots$ span atoms of the cell and indices $\alpha, \beta,
\gamma, \cdots$ are the Cartesian directions of the displacement.
Second order harmonic \IFC, $\Phi(i,j)^{\alpha \beta}$,
calculations were already implemented in the original harmonic \APL\ library
\cite{aflowPAPER} which obtains dispersion curves using three different approaches:
direct force constant  \cite{Maradudin1971,Madelung,Kresse1995}, 
linear response and projector-augmented wave potentials \cite{Gajdos2006}, 
and the frozen phonon methods \cite{Stokes_FROZSL_Ferroelectrics_1995, Boyer_Stokes_Mehl_Ferroelectrics_1995}. 
  
Third order \IFCs, $\Phi(i,j,k)_{\alpha \beta \gamma}$ contain information about the anharmonicity of the lattice 
and they tend to rule phonon scattering in single crystals in the medium-temperature ranges \cite{Feng_prb_2016,Lindsay_nmt_2016}.
Given the choice of a supercell size, the finite difference method
to calculate the third-order \IFCs\ leads to:
\begin{equation}
  \begin{split}
    &\Phi(i,j,k)_{\alpha \beta \gamma} \equiv   \frac{\partial^3V}{\partial r(i)^{\alpha} \partial r(j)^{\beta} \partial r(k)^{\gamma}} \simeq \\
    &\simeq \frac{1}{2h} \left[\frac{\partial^2V}{ \partial r(j)^{\beta} \partial r(k)^{\gamma}}\left(h(i)^{\alpha}\right)-\frac{\partial^2V}{ \partial  h(j)^{\beta} \partial  r(k)^{\gamma}}\left(-h(i)^{\alpha}\right)\right] \\
    &\simeq \frac{1}{4h^2} \bigg[     -\psi\left(h(i)^{\alpha},h(j)^{\beta},k\right)_{\gamma}+\psi\left(-h(i)^{\alpha},h(j)^{\beta},k\right) _{\gamma}\\
    &\ \ \ +\psi\left(h(i)^{\alpha},-h(j)^{\beta},k\right) _{\gamma}-\psi\left(-h(i)^{\alpha},-h(j)^{\beta},k\right) _{\gamma}\bigg]
 \label{eq:3ifc}
 \end{split}
 \end{equation}
where $\left\{\pm h(i)^{\alpha}\right\}, \left\{\pm h(j)^{\beta}\right\}$ represent displacements of magnitude $h$ of
the $i,j$-atoms in the Cartesian directions $\pm\alpha,\pm\beta$ 
and $\psi\left(\pm h(i)^{\alpha},\pm h(j)^{\beta},k\right)_{\gamma}$ are the $\gamma$-components of the forces felt by the
$k$-atom in the distorted configurations caused by the $i$- and $j$-atoms.

The third order \IFCs' calculation is computationally intensive: 
each $\Phi(i,j,k)_{\alpha \beta \gamma}$ requires four supercell calculations (Eq. \ref{eq:3ifc}).
Effective use of crystal symmetry can help the process \cite{Esfarjani_prb_2011}. 
\AAPL\ uses point, factor and space group symmetry operations computed by the
\AFLOW\ symmetry engine \cite{curtarolo:art121} to identify equivalence between
single, pairs and triplets of atoms (positions) and test equivalence between other
field quantities, such as differential $\Phi$ or finite difference forces $\psi$ (covariantly transforming).
To avoid confusion, here we use indices as super
$^{\{\alpha \beta \gamma \cdots \}}$ or sub-scripts $_{\{\alpha \beta  \gamma \cdots \}}$ 
to identify the character of the symmetry transformation to be applied \cite{Lovelock_tensors_1975}.

The reduction of third order \IFC' calculations is performed through the following steps: 

{\bf 1)}
Inequivalent atoms, pairs and triplets are identified using space group symmetries.
The user chooses the neighbor-shell cut-off and only pairs/triplets completely contained are considered. 

{\bf 2)}
The \IFC\ tensors belonging to inequivalent triplets are analyzed. 
The symmetry operations mapping the representative inequivalent to the equivalent $\Phi$ are saved:
$\Phi(i,j,k)_{\alpha \beta \gamma}\rightarrow \Phi'(i',j',k')_{\alpha' \beta' \gamma'}$. 

{\bf 3)}
Each inequivalent tensor $\Phi(i,j,k)_{\alpha \beta \gamma}$ contains $3\times3\times3=27$ coefficients.
Every static {\it ab-initio} calculation produces the vectorial force field for all the $k$-atoms of the supercell 
(where each $k$, combined with the inequivalent pair $(i,j)$, possibly generates $(i,j,k)$ inequivalent triplets)
starting from a combination of deformed positions for the $i-$ and $j$-atoms belonging to inequivalent pairs.
This requires the evaluation of $3\times3=9$ configurations. 
Following Eq. (\ref{eq:3ifc}) four forces $\psi\left(\pm h(i)^{\alpha},\pm h(j)^{\beta},k\right)_{\gamma}$ 
are required for every entry $\Phi(i,j,k)_{\alpha \beta \gamma}$.
To conclude, a total of 36 static calculations are necessary to parameterize   $\Phi(i,j,\forall k\in {\rm sc})$. 

{\bf 4)} 
A large look-up table of all the necessary finite difference forces $\psi\left(h(i)^{\alpha},h(j)^{\beta}, k \right)$ 
is prepared at the beginning of the process. 
Every $\psi$ can be constituent of many inequivalent $\Phi(i,j,k)_{\alpha \beta \gamma}$, 
and, within each $\Phi$, be a term in several internal coefficients. 
To exploit redundancy, the force field generated by every static {\it ab-initio} calculation is mapped through symmetry operations to 
recover as many possible other
$\psi(h(i)^{\alpha},h(j)^{\beta},\forall k\!\!\in {\rm sc})_{\gamma}\rightarrow \psi(h(i')^{\alpha'},h(j')^{\beta'},\forall k'\!\!\in {\rm sc})_{\gamma'}$.
Calculated and symmetry reproduced  $\psi$ are then removed from the table, and the algorithm moves to the next one to characterize.
The process is repeated until all the $\psi$ are found. The process guarantees that only the minimum amount of calculations are performed,
compatible with the model of Eq. (\ref{eq:3ifc}).

{\bf 5)} 
During the process, many equivalent entries of the tensors $\Phi(i,j,k)_{\alpha \beta \gamma}$ are generated by the static {\it ab-initio} calculations.
Because of unavoidable numerical noise, often equivalent entries have slightly different values, and the final value needs to be symmetrized somehow.
This is performed during the re-symmetrization necessary to address the ``sum rules'' conservation.

{\bf Sum rules and re-symmetrization.}
Invariance with respect to any global rigid displacement translates into ``sum rules'' for anharmonic \IFCs:
\begin{equation}
  \sum\limits_{k} \Phi(i,j,k)_{\alpha \beta \gamma} = 0,\,\,\, \forall {\rm \ permutations\ of\ } i,j,k.
  \label{eq:sum}
\end{equation}
Due to finite size effects, the calculated \IFCs\ are not perfectly symmetric and
do not strictly satisfy Eqns. (\ref{eq:sum}), 
causing numerical instabilities.
To tackle the issue, we implement an iterative algorithm which
corrects $\Phi(i,j,k)$ and fulfills the constraints.

Given a set of $\Phi(i,j,k)$ the error $x$ of each sum rule at step
$\mathcal{N}$ is defined as:
\begin{equation}
 x(i,j) ^\mathcal{N}_{\alpha \beta \gamma}\equiv  \sum\limits_{k} \Phi(i,j,k) ^\mathcal{N}_{\alpha \beta \gamma} 
\end{equation}
Each iteration is composed of correction and re-symmetrization of equivalent \IFCs.
Correction, $\Phi(i,j,k)^{\mathcal{N}}_{\alpha  \beta \gamma} \rightarrow \Phi(i,j,k)^{\mathcal{N}+1}_{\alpha \beta  \gamma} $ is
given by:
\begin{equation}
  \begin{split}
    &\Phi(i,j,k)_{\alpha \beta \gamma}^{\mathcal{N}+1}
    =\left(1-\mathcal{\mu}\right)\Phi(i,j,k)_{\alpha\beta\gamma}^{\mathcal{N}}+\frac{\mathcal{\mu}}{n_{\mathrm{eq}}}\times\\
    &\times\!\!\!\!\sum\limits_{\substack{i' j'k'\\\alpha'\beta'\gamma'}}\limits^{\mathrm{eq}}\!\!\left(\rule{0cm}{0.9cm}\right.\!\!\Phi(i'\!,j'\!\!,k')_{\alpha'\beta'\gamma'}^{\mathcal{N}}\!-\!
    \frac{x(i'\!,j')^\mathcal{N}_{\alpha'\beta'\gamma'}\left|\Phi(i'\!,j'\!,k')_{\alpha'\beta'\gamma'}^{\mathcal{N}}\right|}{\sum\limits_{k''}\left|\Phi(i'\!,j'\!\!,k'')_{\alpha'\beta'\gamma'}^{\mathcal{N}}\right|}\left.\rule{0em}{0.9cm}\right)\!,
  \end{split}
  \label{eq:correction}
\end{equation}
where the term $x^{\mathcal{N}}\left|\Phi^{\mathcal{N}}\right|/\sum\left|\Phi^{\mathcal{N}}\right|$
corrects $\Phi$ based on the total error times the absolute contribution of $\Phi$ in the ``sum rule''.
The sum over the combination of indices
$\left\{i',j',k',\alpha',\beta',\gamma'\right\}$ giving \IFCs\
equivalent to $\Phi(i,j,k)_{\alpha \beta \gamma}$ (there are
$n_{\mathrm{eq}}$) is meant to symmetrize the error across all the entries.
The mixing fraction in the iterative process, $\mu$, can be adjusted by the user to optimize convergence rate and robustness.
Overall, with increasing  number of neighbor shells, the user can effectively reduce this systematic error and achieve effective convergence of $\kappa_l$.
\ \\

{\bf Calculation workflows.}

$\bullet$ Anharmonic scattering time $\tau^{\mathrm{anh}}_{\lambda}$:
\begin{equation}
\begin{split}
  &\xrightarrow[{\rm finite\ forces}]{\rm AFLOW-\AAPL} \psi
  \xrightarrow[{\rm force\ constants}]{{\rm Eq.} (\ref{eq:3ifc})} \Phi^{\mathcal{N}} 
  \xrightarrow[{\rm symmetrization}]{{\rm Eq.} (\ref{eq:correction})} \Phi \rightarrow \\
  &\xrightarrow[{\rm scatt.\ matrix}]{{\rm Eq.} (\ref{eq:scattering_V})} V^{\pm} 
  \xrightarrow[{\rm scatt.\ rates}]{{\rm Eqns.} (\ref{gamma2eq}-\ref{gamma3eq})} \Gamma^{\pm}
  \xrightarrow[{\rm anh.\ scatt.\ time}]{{\rm Eq.} (\ref{eq:tau_anh})} \tau^{\mathrm{anh}}_{\lambda}.
  \label{workflow:tau_anh}
\end{split}
\end{equation}

$\bullet$ Elastic scattering time $\tau^{\mathrm{iso}}_{\lambda}$
(isotopic disorder) and grain boundaries scattering time $\tau^{\mathrm{bnd}}_{\lambda}$ (polycrystalline materials):
\begin{equation}
  \begin{split}
    &\xrightarrow[{ab\ initio}]{\rm AFLOW-\APL} \psi
    \xrightarrow[{\rm force\ const,}]{{\rm Eq.} (\ref{eq:potential})} \Phi 
    \xrightarrow[{\rm dynamical\ mat.}]{{\rm Eq.} (\ref{eq:dynamical2})} D(\mathbf{q}) \rightarrow \\
    & \rightarrow 
    \begin{cases}
      \xrightarrow[{\rm phonons}]{{\rm Eq.} (\ref{eq:dynamical1})}  \omega_{\lambda}     
      \xrightarrow[{\rm elastic\ scatt.\ time,\ rate}]{{\rm Eq.}
        (\ref{eq:tau_iso})} \tau^{\mathrm{iso}}_{\lambda},
      \Gamma_{\lambda\lambda'}. \\
      \xrightarrow[{\rm group\ velocities}]{{\rm Eq.} (\ref{eq:group_velocities})} v_{\lambda} 
      \xrightarrow[{\rm grain\ bound.\ scatt.\ time}]{{\rm Eq.} (\ref{eq:tau_bnd})} \tau^{\mathrm{bnd}}_{\lambda}.
    \end{cases} 
    \label{workflow:tau_iso_bnd}  
  \end{split}
\end{equation}

$\bullet$ Conductivity $\kappa^{\alpha\beta}_{l}$:
\begin{equation}
  \begin{split}
    & \left\{ \tau^{\mathrm{anh}}_{\lambda},\tau^{\mathrm{iso}}_{\lambda},\tau^{\mathrm{bnd}}_{\lambda} \right\} \xrightarrow[{\rm total\ scatt.\ time}]{{\rm Eq.} (\ref{eq:tau_zero})} \tau^0_{\lambda} \rightarrow \\
    &\xrightarrow[{\rm mean\ free\ disp.}]{{\rm Eq.} (\ref{equationF})}  \mathbf{F}_{\lambda} \xrightarrow[{\rm conductivity}]{{\rm Eq.} (\ref{equationKappa})} \kappa^{\alpha\beta}_{l}.
    \label{workflow:kappa}
  \end{split}
\end{equation}

\section{Computational details}

\begin{figure*}
  \includegraphics[width=0.98\textwidth]{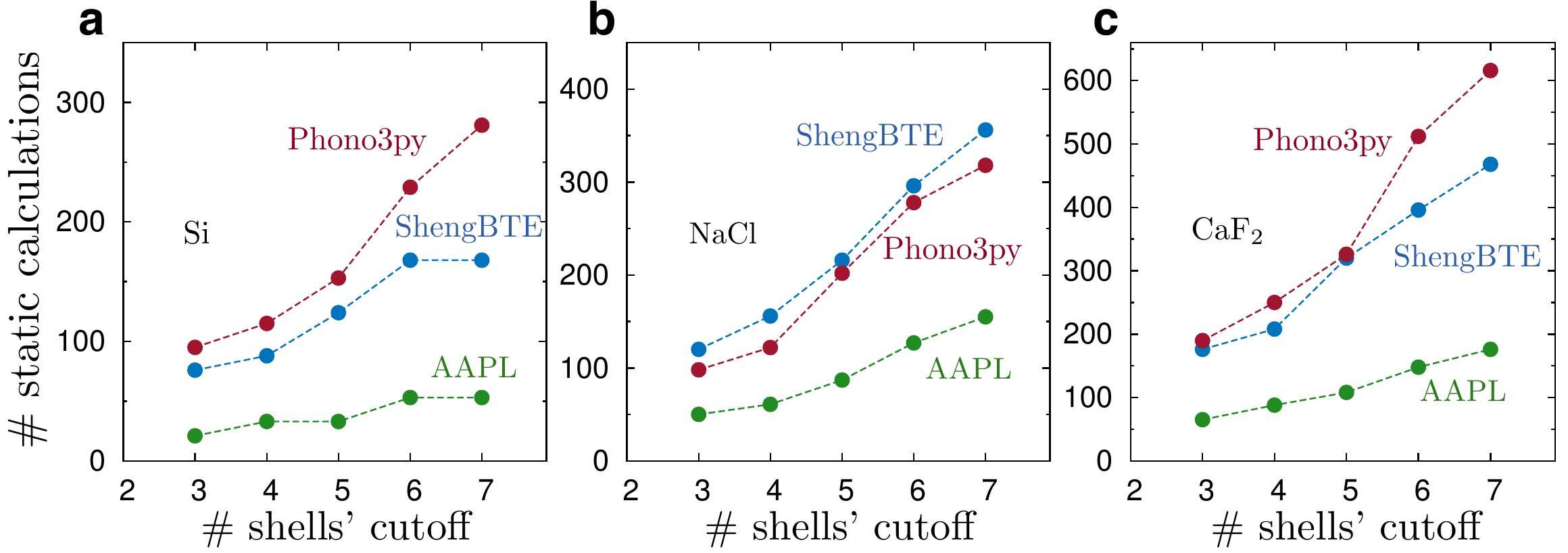}
  \vspace{-3mm}
  \caption{\small 
    Number of required static calculations for {\bf (a)} \ce{Si}, {\bf (b)} \ce{NaCl}, and {\bf (c)} \ce{CaF2} 
    for the computation of the 3rd order \IFCs, applying different cut-offs ($n^{\mathrm{th}}$ neighbor) 
    using \AAPL\ (green), Phono3py (red), and ShengBTE (blue).}
  \label{fig:scaling}
\end{figure*}

{\bf Geometry optimization.} 
All structures are fully relaxed using the automated framework
\AFLOW\ \citeAFLOW\ and the
\VASP\ package \cite{kresse_vasp}. 
Optimizations are performed following the \AFLOW\ standards \cite{Calderon_cms_2015}. 
We use the projector augmented wave ({\small PAW}) potentials \cite{PAW}
and the exchange and correlation functionals parameterized by the generalized gradient approximation proposed
by Perdew-Burke-Ernzerhof (\PBE) \cite{PBE}. 
All calculations use a high energy-cut-off, which is 40$\%$ larger
than the maximum recommended cut-off among all component potentials, and a {\bf k}-point mesh of 8,000 {\bf k}-points per reciprocal atom.
Primitive cells are fully relaxed (lattice parameters and ionic positions) until the energy difference between two consecutive ionic steps
is smaller than $10^{-4}$ eV and forces in each atom are below $10^{-3}$ eV/\AA.

{\bf Phonon calculations.}
Phonon calculations are performed out using the automatic phonon library, \APL, as implemented in \AFLOW, 
and by using \VASP\ to obtain the 2$^{\mathrm{nd}}$ order \IFCs\ via the finite-displacement approach \cite{Pinku_sr_2016}.
The magnitude of the displacement is 0.015 \AA. 
Electronic \underline{s}elf \underline{c}onsistent \underline{f}ield
({\small SCF}) iterations
for static calculations are stopped when the difference of
energy between the last two steps is less than $10^{-5}$ meV. 
The threshold ensures a good convergence for the wavefunction and sufficiently accurate values for forces and harmonic constants.
Non-analytic contributions to the dynamical matrix are also included using the formulation developed by Wang \textit{et al.} \cite{Wang2010}.
Frequencies and other related phonon properties are calculated on a $21\times21\times21$ {\bf q}-point mesh in the Brillouin zone, which is a tradeoff 
between the computational cost,
convergence of  the phonon density of states, pDOS, and the derived thermodynamic properties.
Integrations within the Brillouin zone are obtained by using the linear interpolation tetrahedron method available in \AFLOW.

{\bf Lattice thermal conductivity.}
Anharmonic force constants are extracted from a $4\times4\times4$ supercell using a cut-off that includes all 4$^{\mathrm{th}}$ neighbor shells. 
Thermal conductivity is evaluated on a $21\times21\times21$ {\bf q}-point mesh 
using $\zeta=0.1$ for the Gaussian smoothing, Eq. (\ref{delta2gauss}).
The dense mesh ensures the convergence of the values obtained for $\kappa_{l}$ \cite{ShengBTE_2014}. 
The \IFCs' calculations are iterated self-consistently until all sum rules are smaller $10^{-7}$ eV/\AA$^3$. 

{\bf Analysis of Results.}
Different statistical parameters are used to measure the qualitative and quantitative agreement of \AAPL\ with respect to experimental values.
The Pearson correlation coefficient $r\left[\left\{X\right\},\left\{Y\right\}\right]$ is a measure of the linear correlation between two variables, $\left\{X\right\}$ and $\left\{Y\right\}$:
\begin{equation} 
  \label{Pearson}
  r = \frac{\sum\limits_i \left(X_i - \overline{X} \right) \left(Y_i - \overline{Y} \right) }{ \sqrt{\sum\limits_i \left(X_i - \overline{X} \right)^2} \sqrt{\sum\limits_i \left(Y_i - \overline{Y} \right)^2}},
\end{equation}
where $\overline{X}$ and $\overline{Y}$ are the averages of $\left\{X\right\}$ and $\left\{Y\right\}$.

The Spearman rank correlation coefficient $\rho\left[\left\{X\right\},\left\{Y\right\}\right]$ is a measure of the monotonicity of the relationship between two variables.
The values of the two variables  $\left\{X\right\}$ and $\left\{Y\right\}$ are sorted in ascending order, and are assigned rank values  $\left\{x\right\}$ and $\left\{y\right\}$ which
are equal to their position in the sorted list. The correlation coefficient is then given by
\begin{equation} 
  \label{Spearman}
  \rho = \frac{\sum\limits_i \left(x_i - \overline{x} \right) \left(y_i - \overline{y} \right) }{ \sqrt{\sum\limits_i \left(x_i - \overline{x} \right)^2} \sqrt{\sum\limits_i \left(y_i - \overline{y} \right)^2}}.
\end{equation}
$\rho$ is useful for determining how well the values of one variable can predict the ranking of the other variable.

We also investigate the \underline{r}oot-\underline{m}ean-\underline{s}quare \underline{r}elative \underline{d}eviation, \RMSRD, of the calculated $\kappa$ {\it versus} the experiment. 
The \RMSRD\ will measure the quantitative difference between  \AAPL\ and experimental results:
\begin{equation} 
  \label{RMSD}
  {\rm \small RMSrD} = \sqrt{\frac{ \sum\limits_i\ \left( \frac{X_i - Y_i}{X_i} \right)^2 }{N_{\{X,Y\}} - 1}} ,
\end{equation}
Lower values of \RMSRD\ indicate better agreement.

\section{Results}

{\bf Scaling.}
The calculation of the anharmonic \IFCs\ is the most computationally expensive step in the method. 
First, we test the number of required calculations for some structural prototypes such as 
diamond (spacegroup: $Fd\overline{3}m,\ \#227$; Pearson symbol: $cF8$; \textit{Strukturbericht} designation: A4; \AFLOW\ Prototype: {\sf A\_cF8\_227\_a} \cite{curtarolo:art121}), 
rocksalt ($Fm\overline{3}m,\ \#225$, $cF8$, B1, {\sf AB\_cF8\_225\_a\_b} \cite{curtarolo:art121}) and 
fluorite ($Fm\overline{3}m,\ \#225$, $cF12$, C1, {\sf AB2\_cF12\_225\_a\_c} \cite{curtarolo:art121}) 
for which there are abundant available experimental data. 
We compare the number of calculations
and how they scale with respect to  the chosen cut-off for the \IFCs\ (see Figure \ref{fig:scaling}) 
for different available software (Phono3py and ShengBTE software packages).
The number of required static calculations increases with the cell's complexity, the total number of atoms, 
and the number of inequivalent positions in the primitive cell. 
\AAPL\ reduces the number of required calculations compared to the other two codes for the three tested prototypes,
indicating that the \AAPL\ algorithm is efficient at handling symmetry equivalence.
For example, in silicon and using the minimum shell cut-off, \AAPL\ only needs 21 calculations, while ShengBTE requires 76.
The advantage is preserved while increasing the range of the interactions.
For example, Phono3py requires 616 static calculations for \ce{CaF2} with 7$^{\mathrm{th}}$ neighbor shells, 
whereas \AAPL\ needs less than one third of this amount (176). Figure \ref{fig:scaling} summarizes the scaling results.

\begin{figure}[h]
  \includegraphics[width=0.98\columnwidth]{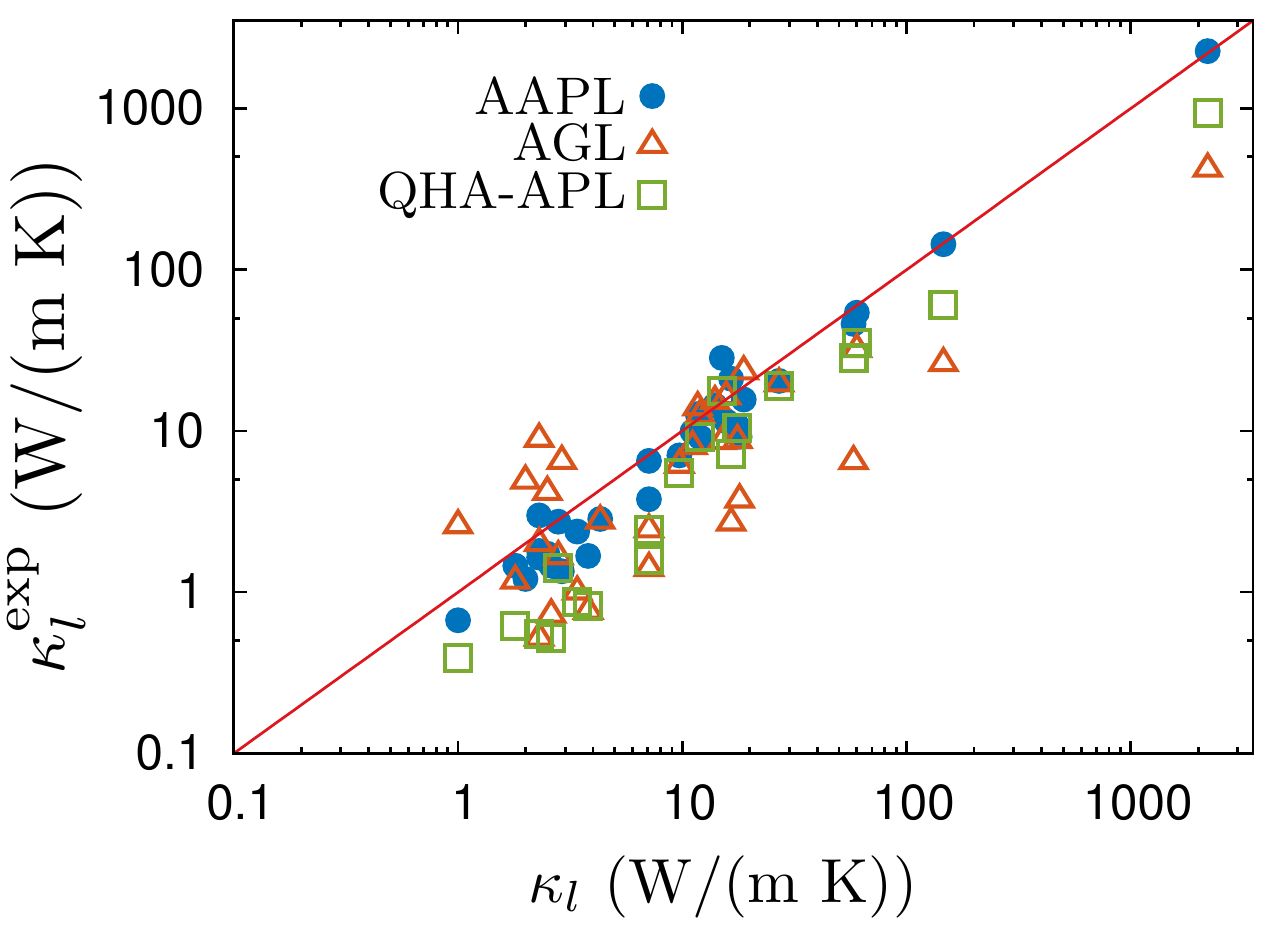}
  \vspace{-5mm}
  \caption{\small
    Calculated lattice thermal conductivities at 300 K {\it versus} experimental. 
    Blue circles are used for \AAPL\ results, 
    empty orange triangles for the quick \AFLOW-\AGL\ prediction of Refs. \cite{curtarolo:art96, aflowlibPAPER},
    and empty green squares for \AFLOW-\QHAAPL\ results of Ref. \cite{Pinku_sm_2016}. 
    The red line represents equality (calculation=experiments).}
  \label{fig:validation}
\end{figure}

{\bf Validation with experiments.}
A data set of 30 compounds is used to validate our framework.
The list of materials includes semiconductors and insulators that belong to different
structural prototypes such as 
diamond ({\sf A\_cF8\_227\_a} \cite{curtarolo:art121}), 
rocksalt ({\sf AB\_cF8\_225\_a\_b} \cite{curtarolo:art121}), and 
fluorite ({\sf AB2\_cF12\_225\_a\_c} \cite{curtarolo:art121}).
To maximize the heterogeneity of the data set, we select materials containing
as many different elements as possible from the \textit{s}-, \textit{p}-, and \textit{d}-blocks of
the periodic table.
The comparison of calculated {\it versus} experimental values of $\kappa_{l}$ is 
summarized in Table \ref{tab:validation} and Figure \ref{fig:validation}.

\begin{table}[]
  \caption{\small
    Calculated and experimental lattice thermal conductivity of
    diamond (Strukturbericht: A4; \AFLOW\ standardized prototype name {\sf A\_cF8\_227\_a}
    \cite{curtarolo:art121}),
    rocksalt (B1, {\sf AB\_cF8\_225\_a\_b} \cite{curtarolo:art121}) and 
    fluorite (C1, {\sf AB2\_cF12\_225\_a\_c} \cite{curtarolo:art121}) structure semiconductors and insulators at 300 K. Units: $\kappa_{l}$ in W/(m K).}
  \begin{center}
    \begin{tabular}[b]{lcccrr}
      \hline \hline
      Formula & Pearson & s.g.\# & {\small Struk.}$^a$ & $\kappa_{l}^{\sAAPL}$ & $\kappa_{l}^{\sexp}$ \\
      \hline
      \ce{C} & $cF8$ & 227 & A4 & 2270 & 2200 \cite{Anthony_prb_1990,Wei_prl_1993} \\ 
      \ce{Si} & $cF8$ & 227 & A4 & 144 & 146 \cite{Touloukian_Serie}\\
      \ce{Ge} & $cF8$ & 227 & A4 & 46.17 & 58 \cite{Carruthers_prsa_1957} \\
      \ce{AgCl} & $cF8$ & 225 & B1 & 0.67 & 1 \cite{Morelli_Slack_2006}         \\      
      \ce{BaO} & $cF8$ & 225 & B1 & 2.99 & 2.3 \cite{Morelli_Slack_2006}        \\
      \ce{CaO} & $cF8$ & 225 & B1 & 20.41 & 27 \cite{Morelli_Slack_2006}        \\
      \ce{KBr} & $cF8$ & 225 & B1 & 2.38 & 3.4 \cite{Morelli_Slack_2006} \\ 
      \ce{KCl} & $cF8$ & 225 & B1 & 3.77 & 7.1 \cite{Morelli_Slack_2006} \\
      \ce{KI} & $cF8$ & 225 & B1 & 1.46 & 2.6 \cite{Morelli_Slack_2006} \\
      \ce{LiF} & $cF8$ & 225 & B1 & 10.73 & 17.6 \cite{Morelli_Slack_2006} \\
      \ce{LiH} & $cF8$ & 225 & B1 & 28.4 & 15 \cite{Morelli_Slack_2006} \\
      \ce{MgO} & $cF8$ & 225 & B1 & 54.06 & 60 \cite{Morelli_Slack_2006} \\
      \ce{NaBr} & $cF8$ & 225 & B1 & 2.74 & 2.8 \cite{Morelli_Slack_2006} \\
      \ce{NaCl} & $cF8$ & 225 & B1 & 6.53 & 7.1 \cite{Morelli_Slack_2006} \\
      \ce{NaF} & $cF8$ & 225 & B1 & 21.11 & 16.5 \cite{Morelli_Slack_2006} \\
      \ce{NaI} & $cF8$ & 225 & B1 & 1.46 & 1.8 \cite{Morelli_Slack_2006} \\
      \ce{PbS} & $cF8$ & 225 & B1 & 1.35 & 2.9 \cite{Morelli_Slack_2006} \\
      \ce{PbSe} & $cF8$ & 225 & B1 & 1.21 & 2.0 \cite{Morelli_Slack_2006} \\
      \ce{PbTe} & $cF8$ & 225 & B1 & 1.73 & 2.5 \cite{Morelli_Slack_2006} \\ 
      \ce{RbBr} & $cF8$ & 225 & B1 & 1.68 & 3.8 \cite{Morelli_Slack_2006} \\
      \ce{RbI} & $cF8$ & 225 & B1 & 1.64 & 2.3 \cite{Morelli_Slack_2006} \\
      \ce{SrO} & $cF8$ & 225 & B1 & 9.12 & 12 \cite{Morelli_Slack_2006} \\
      \ce{CdF2} & $cF12$ & 225 & C1 & 3.01 & 4.30 \cite{Popov_pss_2010} \\
      \ce{SrCl2}& $cF12$ & 225 & C1 & 1.80 & 2.3 \cite{Moore1985} \\
      \ce{Mg2Si}& $cF12$ & 225 & C1 & 15.67 & 18.8$^b$ \cite{Martin_jap_1974} \\ 
      \ce{Mg2Ge}& $cF12$ & 225 & C1 & 11.49 & 15.7$^b$ \cite{Martin_jap_1974} \\
      \ce{Mg2Sn}& $cF12$ & 225 & C1 & 9.91 & 11.1$^b$ \cite{Martin_jap_1974} \\
      \ce{Mg2Pb}& $cF12$ & 225 & C1 & 9.20 & 18$^b$ \cite{Martin_jap_1974} \\
      \ce{CaF2} & $cF12$ & 225 & C1 & 7.04 & 9.76 \cite{Popov_pss_2010} \\
      \ce{CeO2} & $cF12$ & 225 & C1 & 11.35 & 10.8 \cite{Jha_RE} \\
      \ce{ThO2} & $cF12$ & 225 & C1 & 14.42 & 14 \cite{Mann_cgd_2010} \\ 
      \hline \hline
      {\footnotesize $^a$ \textit{Strukurbericht}  } \\
      {\footnotesize $^b$ $\kappa_{l}$ at 200 K }
    \end{tabular}
    \label{tab:validation}
  \end{center}
\end{table}

We use different statistical quantities to measure qualitative and
quantitative agreement between the \AAPL\ and experimental results (Table \ref{tab:stats}).
\AAPL\ results strongly correlate with experimental findings, with relatively small \RMSRD\ from experiment 
demonstrating the reliability and robustness of the framework. 
The algorithm should not be blamed for systematic errors in the {\it ab-initio} characterization of the compounds (such as the ones containing Pb). 

We also compare \AAPL\ with approximate phenomenological frameworks
such 
as \AFLOW-\AGL\ \cite{curtarolo:art96} and \AFLOW-\QHAAPL\ \cite{Pinku_sr_2016, Pinku_sm_2016}.
Qualitatively,
all three frameworks have high linear correlation with experiments (Pearson, $r$);
\AAPL\ and \QHAAPL\ are also very effective in rank ordering the compounds (Spearman, $\rho$).
Quantitatively,  \AAPL\ has the lowest \RMSRD\ value, followed by \QHAAPL\ and \AGL.
Accuracy strongly correlates with computational costs (\AAPL$\gg$\QHAAPL$>$\AGL),
so that the users can choose which technique best fulfills their screening needs.

\begin{figure*}[]
  \begin{center}\includegraphics[width=0.98\textwidth]{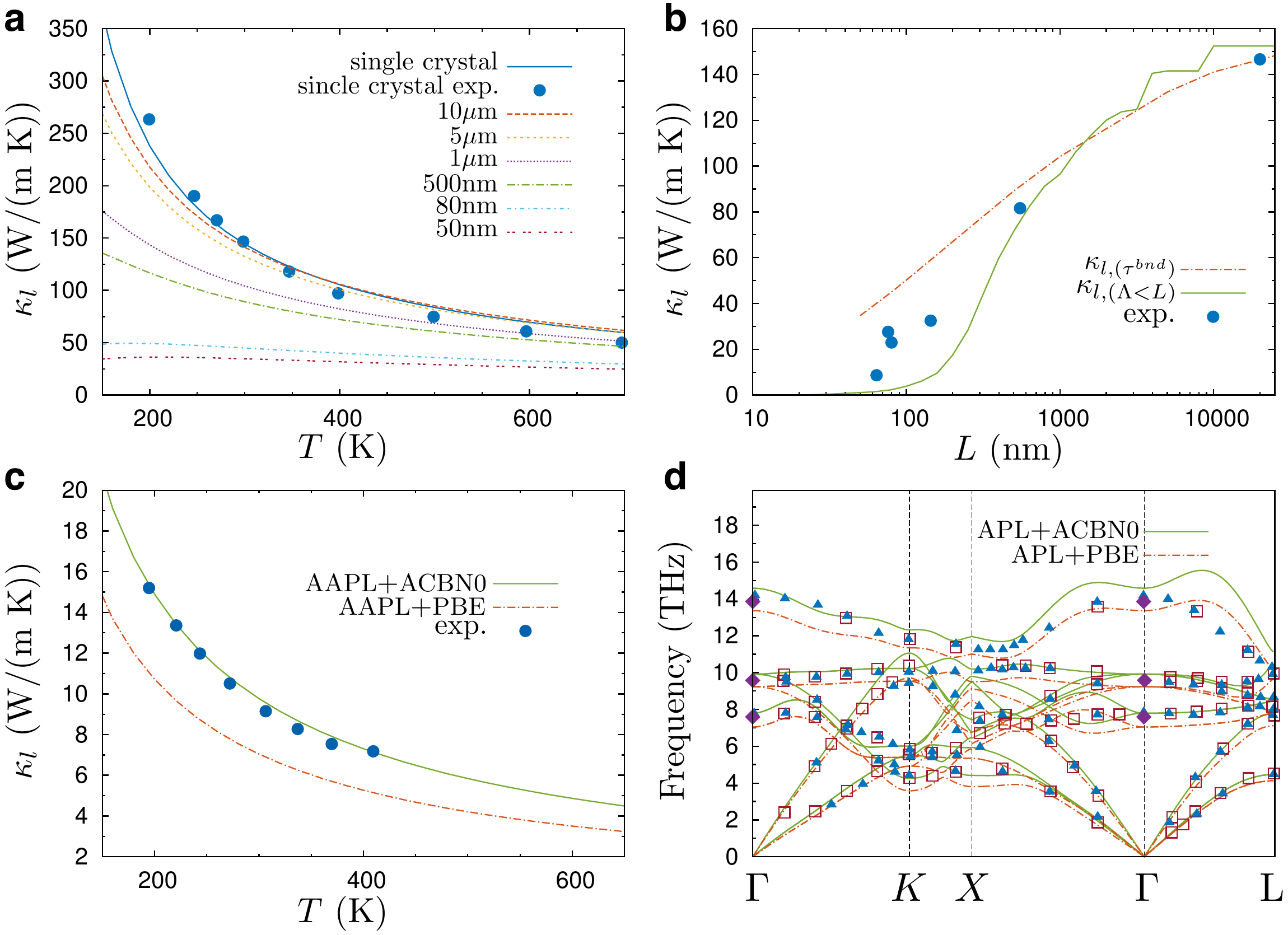}\end{center}
  \vspace{-4mm}
  \caption{\small
    {\bf (a)} Calculated lattice thermal conductivity for single-crystal (blue) and nanocrystalline silicon with different grain size. 
    Blue circles represent measurements for single-crystal Si from Ref. \cite{Touloukian_Serie}.
    {\bf (b)} Cumulative lattice thermal conductivity, $\kappa_{l,(\Lambda<L)}$, (green) of Si as a function of the average grain size, $L$ , at 300 K.
    Lattice thermal conductivity (orange) including the scattering of phonons due to grain boundaries (see Eq. \ref{eq:tau_bnd}) is also presented.
    Blue circles represent experimental data from Ref. \cite{Wang_nl_2011}.
    {\bf (c)} Lattice thermal conductivity of \ce{CaF2} within the
    \ACBN0\ method (green) and \PBE\ functional (orange). 
    Blue circles represent experimental data from Ref. \cite{Slack_PR_1961}.
    {\bf (d)} Phonon dispersion of \ce{CaF2} within the \ACBN0\ method
    (green). The \PBE\ phonon dispersion (orange) is also shown for comparison. 
    Blue triangles and open squares represent neutron scattering data
    from Ref. \cite{Schmalzl_prb_2003} and Ref. \cite{Elcombe_jpcss_1970} respectively.
    Purple diamonds represent Raman and infrared data from Ref. \cite{Kaiser_pr_1962}.
  }
  \label{fig:Si_CaF2}
\end{figure*}

\
\begin{table}
  \caption{\small 
    Root mean square relative deviation (\RMSRD), and Pearson and Spearman correlation for the material data set.}
  \centering
  \begin{tabular}[b]{c C{1.2cm} C{1.6cm} C{1.2cm} }
    \hline \hline
    & \AGL\ & \QHAAPL\ & \AAPL\ \\
    \hline
    $r$ & 0.997 & 0.999 & 0.999 \\
    $\rho$ & 0.706 & 0.976 & 0.933 \\
    RMSrD & 86.20\% & 58.16\% & 27.57\% \\
    \hline \hline
  \end{tabular}
  \label{tab:stats}
\end{table}

{\bf Single-crystal and nanocrystalline silicon.}
Silicon is the perfect benchmark for testing the reliability of \AAPL:
extensive availability of experimental data for well characterized
samples \cite{Wang_nl_2011, Touloukian_Serie} and limited
computational cost due to the diamond
crystal structure with two atoms in the primitive cell and fcc lattice.
Figure \ref{fig:Si_CaF2}(a) depicts the calculated lattice thermal
conductivity at different temperatures for single-crystal and
polycrystalline samples compared to single-crystal experimental values
from Ref. \cite{Touloukian_Serie}.
Boundary effects can be included in two ways:
{\bf i.} by calculating the complete $\kappa^{\alpha\beta}_{l}(L)$ (workflow (\ref{workflow:kappa})) for average grain sizes having different $\tau^{\mathrm{bnd}}_{\lambda}(L)$ (Workflow (\ref{workflow:tau_iso_bnd})) 
or 
{\bf ii.} by neglecting  $\tau^{\mathrm{bnd}}$ from the total scattering time $\tau^0_{\lambda}$ (Eq. (\ref{eq:tau_zero})) and restricting the summation to the modes with a mean free path shorter than $L$
($\Lambda<L$, Eq. (\ref{eq:cumulative_kappa})): $\kappa^{\alpha\beta}_{l,(\Lambda<L)}$.
Both approaches are implemented in \AAPL. Comparison with experimental values for 
different polycrystalline Si samples (average grain size $L$=64, 76, 80, 155, 550 and 20000 nm) at 300 K are presented in 
Figure \ref{fig:Si_CaF2}(b).
Both approximations of grain boundary scattering effects show the same trend and are very close to the experimental results, corroborating the validity of our approaches.

{\bf Extension to \ACBN0\ pseudo-hybrid functional.}
The accuracy of the results ultimately relies on the quality of the
computed \IFCs\ with {\it ab-initio}. 
The use of hybrid functionals \cite{HSE} or advanced electronic
structure methods such as  $GW$ \cite{Kresse_GW} to compute the \IFCs\ 
is limited \cite{forces_GW_Mauri_PRB_2008,Hummer_prb_2009} because of their computational costs. 
Recently, the \ACBN0\ functional was introduced in order to facilitate the accurate characterization of
electronic properties of correlated materials \cite{curtarolo:art93}.
\ACBN0\ is a pseudo-hybrid Hubbard density functional that introduces a
new self-consistent \textit{ab-initio} approach to compute $U$ 
without the need for  empirical parameters.
\ACBN0\ can improve not only the description of the electronic structure, but also the prediction of the 
structural and the vibrational parameters of  solids \cite{curtarolo:art103}.
One of the reasons for this is the better prediction of the Born
charges, $\mathbf{Z}$, and the dielectric tensor, $\epsilon_{\infty}$,
compared to \LDA\ or \GGA\ functionals \cite{curtarolo:art103}.
If the \ACBN0\ functional improves the vibrational parameters of 
solids, we can assume that the calculations of other temperature-dependent properties such as $\kappa_{l}$ may be improved too. 
As a testbed, we chose calcium fluoride, \ce{CaF2},
because of the ample available experimental data \cite{Popov_pss_2010,Schmalzl_prb_2003,Elcombe_jpcss_1970,Kaiser_pr_1962}.
\ce{CaF2} has been extensively used in optical devices due to its low
refractive index, wide band gap, low dispersion, and large broadband
radiation transmittance \cite{Cazorla_prl_2014,Sang_apl_2011,Lyberis_om_2012}.

We use the package \AFLOWPI\ \cite{aflowPI} to obtain the \ACBN0\ electronic
structure of \ce{CaF2}. We obtain $U_{\mathrm{eff}}$ of 13.43 for F-$p$
orbitals. We then use this value inside \AAPL\ for the rest of the
calculations.
\AAPL+\ACBN0\ almost perfectly predicts the experimental $\kappa_{l}$ in
contrast with  \AAPL+\PBE\  which greatly underestimates $\kappa_{l}$ over
the entire temperature range (Figure \ref{fig:Si_CaF2}(c)).
Phonon band structures have also been calculated with the harmonic
library (\APL) to explain the difference (Figure \ref{fig:Si_CaF2}(d)).
\ACBN0\ reproduces the phonon dispersion better than the \PBE\ functional. 
\PBE, as a \GGA\ functional, overestimates bond length and hence it
tends to underestimate vibrational frequencies \cite{Sholl_DFTintro}.
On the contrary, \ACBN0\ describes the bond length more accurately, 
obtaining frequencies higher than \PBE\ and closer to the experimental values \cite{Schmalzl_prb_2003}.
Major differences between \ACBN0\ and \PBE\ results come from the optical bands, so we compared the two main 
properties that are involved in the splitting of the optical band due to the non-analytical contributions
to the dynamical matrix.
While the Born charges are similar for \ACBN0\ and \PBE\ (2.33 e and 2.34 e respectively), 
there are significant differences in the high-frequency dielectric constant $\epsilon_{\infty}$ (2.083 and 2.305 respectively).
The value obtained using  \ACBN0\ is closer to the experimental $\epsilon_{\infty}$ (2.045)  \cite{Kaiser_pr_1962}  than that obtained using PBE.

\section{Conclusions}

The
\underline{A}utomatic-\underline{A}nharmonic-\underline{P}honon-\underline{L}ibrary,
\AAPL, was developed to compute the third order \IFCs\ and solve the \BTE\ within the high-throughput \AFLOW\ framework.
This code automatically predicts the lattice thermal conductivity of
single-crystals and polycrystalline materials using a single input file
and with no further user intervention. 
The symmetry analysis has been optimized to further reduce the number of static calculations compared to other packages.
The robustness and accuracy of the code have been tested with a set of 30 materials that belong to different space groups.
\APL\ has been combined with the \ACBN0\ pseudo-hybrid functional to predict the lattice thermal conductivity of \ce{CaF2}.
Our results demonstrate that using \ACBN0\ can improve not only  the
electronic structure description of the material compared to the \GGA\ functional, but also 
phonon-dependent properties such as the thermal conductivity. 

\section{Acronyms in the \AFLOW\ package}
\AAPL: \underline{A}utomatic-\underline{A}nharmonic-\underline{P}honon-\underline{L}ibrary;
\AGL: {\small \underline{A}FLOW}-\underline{G}ibbs-\underline{L}ibrary  \cite{curtarolo:art96};
\APL:  \underline{A}utomatic-\underline{P}honon-\underline{L}ibrary \cite{curtarolo:art63,aflowPAPER};
\AEL: {\small \underline{A}FLOW}-\underline{E}lastic-\underline{L}ibrary \cite{curtarolo:art115};
\QHA: \underline{q}uasi\underline{h}armonic \underline{a}pproximation \cite{Pinku_sm_2016};
\ACBN0: \underline{A}gapito \underline{C}urtarolo
\underline{B}uongiorno \underline{N}ardelli {\it ab-initio} DFT
functional \cite{curtarolo:art93};
\AFLOWPI:
A minimalist  {\small \underline{A}FLOW}-\underline{Py}thon approach to high-throughput {\it ab initio} calculations
including the generation of tight-binding hamiltonians and the calculation of the \ACBN0\ functional \cite{aflowPI}.

\section{Acknowledgements}
We thank 
Drs. Kristin Persson, Gerbrand Ceder, Natalio Mingo, David Hicks, Mike
Mehl, Ohad Levy, and Corey Oses for various technical discussions.
We acknowledge support by the DOE (DE-AC02-05CH11231), specifically the Basic Energy Sciences program under Grant \# EDCBEE.
C.T.,  M.F., M.B.N. and S.C. acknowledge partial support by DOD-ONR (N00014-13-1-0635, N00014-11-1-0136, N00014-15-1-2863).
The \AFLOW\ consortium acknowledges Duke University -- Center for Materials Genomics --- and the CRAY corporation for computational support.

\newcommand{\Ozolins}{Ozoli\c{n}\v{s}}

\end{document}